\pdfoutput=1
\documentclass[journal=jacsat,manuscript=communication]{achemso}
\usepackage[version=3]{mhchem} 
\usepackage{lineno,xcolor}

\usepackage{graphicx}
\usepackage{subcaption}

\usepackage[english]{babel} 

\usepackage{amsmath}
\usepackage{bm}

\usepackage{siunitx}

\title{Molecular ferroelectric contributions to anomalous hysteresis in hybrid perovskite solar cells}

\author{Jarvist M. Frost}
\author{Keith T. Butler}
\affiliation{Centre for Sustainable Chemical Technologies and Department of Chemistry, University of Bath, Claverton Down, Bath BA2 7AY, UK}

\author{Aron Walsh}
\email{a.walsh@bath.ac.uk}
\affiliation{Centre for Sustainable Chemical Technologies and Department of Chemistry, University of Bath, Claverton Down, Bath BA2 7AY, UK}

\begin{document}
\begin{abstract}
We report a model describing the molecular
orientation disorder in \ce{CH3NH3PbI3}, solving a classical Hamiltonian
parametrised with electronic structure calculations, with the nature of the
motions informed by \textit{ab-initio} molecular dynamics.
We investigate the temperature and static electric field dependence of the 
equilibrium ferroelectric (molecular) domain structure and resulting 
polarisability. A rich domain structure of twinned molecular 
dipoles is observed, strongly varying as a function of temperature and applied electric field. 
We propose that the internal electrical fields associated with microscopic
polarisation domains contribute to hysteretic anomalies in the current--voltage
response of hybrid organic-inorganic perovskite solar cells due to variations
in electron-hole recombination in the bulk.  
\end{abstract}


\maketitle 


Solar cells based on hybrid organic-inorganic perovskites display unusual
device physics, which is still in the process of being
understood.\cite{edri_elucidate_2014,gonzalez_working_2014,snaith_anomalous_2014,even2014analysis,frost-2014} 
One unusual aspect is the notable hysteresis in current--voltage curves,
depending on rate of measurement and starting point within the
curve.\cite{snaith_anomalous_2014} 
When 
the solar cell is held at short--circuit (in the dark or light), the
photovoltaic performance decreases considerably. 
When the photovoltaic cell is operating at near open--circuit voltage,
performance builds. 

The highest apparent efficiencies are produced after 
tempering the solar cell in forward bias at $1.4$ \si{V}.\cite{snaith_anomalous_2014} 
Similar behaviour is observed in both mesoporous and planar solar cell architectures
and independent of the presence of a hole transport layer (e.g. Spiro‐MeOTAD).
This effect is strongest in planar device architectures, as would be expected
if it resulted from the electric field applied between the contacts. 
This behaviour occurs over time scales up to hundreds of seconds.

Perovskite (\ce{ABX3}) structured materials tend to have a large dielectric
constant due to the relative ease of polarising the cell structure.
In particular, distortion of the edge-sharing \ce{BX6} octahedra can produce an
overall electric dipole between the A and B lattice sites.  
Inorganic perovskites are known to exhibit a range of ferroelectric and
ferroelastic phase transitions.\cite{glazer-3384}
The hybrid perovskite analogues are formed by replacing the element at the
icosahedral A site with an isovalent molecule.
In this work we study the photovoltaic absorber
methyl-ammonium lead iodide, \ce{CH3NH3PbI3} (MAPI). 
Here \ce{CH3NH3+} (MA) is a singly-charged closed-shell molecular cation.
As compared to inorganic perovskites, the hybrid materials have a lower crystal
symmetry and the possibility of dipolar cation interactions gives rise to strong
low-frequency dielectric screening and the formation of polarised domain
structures.\cite{frost-2014}

Ferroelectric materials have been studied in the context of photovoltaic
applications for over half a century.\cite{ferroelectrics} A number of effects have
been attributed to the lack of centrosymmetry. The anomalous photovoltaic
effect was first reported in samples of PbS\cite{Starkiewicz1946} in 1946 and was
subsequently linked to the existence of a ferroelectric transition,
resulting in large photovoltages (\textit{ca.} 10 \si{\kilo\volt}).\cite{Fridkin1978} Similar
effects have been reported for ferroelectric phases of SbSI, ZnS and CdTe.\cite{Fridkin1978,Neumark1962,Johnson1975}
Additionally, oxide perovskites such as \ce{BaTiO3} and \ce{KNbO3} demonstrate
a bulk photovoltaic effect.\cite{Baltz1981} 
In these materials photocurrents can be generated in the absence of asymmetric
electrical contacts, unlike standard photovoltaic cells. 

In this Letter we report the implementation of a classical Monte-Carlo
simulation of the domain behaviour arising from molecular rotation in hybrid
perovskites.  
The model takes parameters from density functional theory (DFT)
calculations, using both static lattice and molecular dynamic (MD)
simulations. 
In this work we restrict ourselves to the two-dimensional case, and allow the
dipoles to freely rotate (no cage strain is applied).

We propose that the MA ions in this material are highly
rotationally mobile, the interaction between these ions forms ordered
domains (which respond slowly to applied electric fields), which results in
a structured local potential field. 
We speculate that the equilibrium (open--circuit) configuration is
beneficial for solar cell operation by reducing charge carrier recombination
through interpenetrating percolating pathways of (electric potential) valleys
and ridges for holes and electrons. 
At short--circuit, the electric field resulting from the built-in voltage
is sufficient to disrupt this structure, suppressing long-range order and
resulting in more isolated domains. 
The existence of intricate dipole phase behaviour and the resultant structure
in internal electric fields indicate that these photoferroic characteristics, atypical of
standard photovoltaic materials, must be considered in device modelling. 

\textit{Low-frequency dielectric behaviour.}
The dielectric response of \ce{CH3NH3PbI3}, and related hybrid perovskites, 
exhibits significant temperature and frequency dependence. 
At low temperatures there is a discontinuity  associated with the first-order
phase transition between the orthorhombic and tetragonal phases (\textit{ca.} 160 K); at
higher temperatures the orientation of MA becomes (partially) disordered. 
Work by Poglitsch and Weber in 1987 measured the complex dielectric response of
methyl-ammonium lead halides (iodide, chloride, bromide) as a function of
temperature between $100-300$ \si{\kelvin}. 
\cite{poglitsch_dynamic_1987}
The effective dielectric constant at 300 K was measured to be 33 for \ce{CH3NH3PbI3} at
a frequency of 90 GHz.  
In 1992, Onoda-Yamamuro \textit{et al} reported a value of \textit{ca.} 58 at
a frequency of 1 kHz.\cite{onoda-935}
In contrast, the static dielectric constant, in the absence of molecular reorientation, is predicted to be 24.1 from electronic structure calculations (PBEsol + QS\textit{GW}),\cite{brivio-155204} which is in good agreement with the value of 23.3 determined
from a fit of permittivity measurements over 100--300 K to the Kirkwood-Fr\"{o}hlich equation.\cite{onoda-935}
The unusual dielectric behaviour will make analysis of impedance measurements on
photovoltaic cells challenging.

%
%

\textit{Ab initio molecular dynamics.}
An open question in these materials is the alignment and dynamics of the
MA ion.  
Analysis of $^2$H and $^{14}$N NMR spectra confirmed that MA cation rotation is a rapid
process at room temperature.\cite{wasylishen-581}  
X-ray diffraction has been used to characterise the low temperature orthorhombic ($Pna2_1, C_{2v}$ symmetry), room temperature tetragonal ($I4/mcm, D_{4h}$ symmetry) and above room temperature cubic ($Pm3m, O_h$ symmetry) crystal structures of MAPI.\cite{poglitsch_dynamic_1987}
The position of the MA molecules is usually described with partial occupancies that satisfy the space group symmetry,\cite{poglitsch_dynamic_1987,onoda-1385} e.g. in the cubic phase eight identical positions can be fitted around the standard ($\frac{1}{2},\frac{1}{2},\frac{1}{2}$) perovskite site, each occupied with equal probability.\cite{poglitsch_dynamic_1987} 
There is a distinct first-order orthorhombic-to-tetragonal phase transition, but
the tetragonal-to-cubic transition is close to second-order with no change in pseudo-cubic cell volume.\cite{kawamura-1694} 
It should be noted that in hybrid perovskite thin-films, analysis of X-ray scattering data\cite{choi-127} and electron microscopy\cite{zushi-916} has suggested the presence of a lower symmetry nanostructure, as well as the appearance of the cubic phase at room temperatures. 

To provide atomistic insight, without the assumption or restriction of lattice
symmetry beyond a periodic supercell (80 atom $2\times2\times2$ expansion of the pseudo-cubic perovskite structure\cite{brivio-042111}), we investigate the energetics
and dynamics of MA in MAPI with \textit{ab-initio} MD simulations based on the PBEsol 
exchange-correlation functional. 
We employ a timestep of $0.5$ \si{fs}, which is sufficient to describe even the
C--H vibrations. 
The MD trajectory at 300 K contains large-scale fluctuation of the ions about their
equilibrium positions, including rapid rotation of the methyl group and total
rotation of the methyl-ammonium ion. 
(A video file of the trajectory is provided elsewhere.\cite{MAPIVideoFigshare}) 
The hybrid perovskites are structurally soft materials; so far almost all
published calculations and analysis have assumed perfect crystals, while these data
indicate that such structures are not representative of the materials
at room temperature. 
Further structural analysis is on-going. 

Due to the permanent molecular dipole of the methyl-ammonium ion,\cite{frost-2014} its ensemble
average position and the dynamics of its movements are of interest in
explaining the dielectric response and electrical behaviour of devices made
from MAPI.
Custom codes were written to analyse the MD trajectories, identifying the C--N
bonds across the periodic boundaries and calculating this molecular alignment
(of the eight MA ions) relative to the pseudo-cubic unit cell. 
The distribution of spherical coordinates over the MD ensemble 
enables us to make
statements about the average distribution of molecular direction, relative to
the crystallographic unit cell. 
Here the histogram is in binned in $(\theta,\phi)$ spherical angles, and
hence the bins are not a constant solid angle.
The $O_h$ symmetry of the ideal cubic perovskite phase leads to a 48--fold
reduction of the phase space onto its reflection domain. 

When we plot the data without considering the symmetry of the MA ion environment (Figure
\ref{theta_phi_nosymm}), there is little that we can say other than a preference
for the ion to align with the faces of the cube ($\Delta\theta=\pi/_2$). 
The limited simulation time leads to the evident incomplete coverage of spherical phase space. 
Therefore we reflect the data onto the first octant, and further exploit symmetry to
reduce the internal coordinates to contain the domain between the unique
[1,0,0] (X) faces, [1,1,0] (M) edges and [1,1,1] (R) diagonals (Figure
\ref{theta_phi_symm}) to increase the signal to noise ratio by a factor of 48 
(see Supplementary Material for more details).
This reveals a high density of ensembles in a distribution around
facial alignment, a lowered distribution around edge alignment, and an
increased distribution in a disordered halo around diagonal alignment.

We can further quantify these distributions by binning the ensemble of
symmetry reduced alignment vectors by whether they are nearest (in angle) to
the face, edge or diagonal vectors. 
Doing so we find that the raw densities are $35\%$ face, $42\%$ edge and
$23\%$ diagonally aligned (populations $[6497, 7822, 4228]$). 
Due to the symmetry of these orientations (with 6, 12 and 8 fold degeneracy, respectively), and the boundaries between
these domains, these populations are not directly comparable (the solid angles they cover are different). 
Therefore we weight these distributions with surface areas evaluated from a flat
spherical distribution (calculated with a Monte-Carlo integration of $10^{5}$
points, using the same codes). 
These densities, which would be equal if there was no orientation preference,
are $42\%$ face, $31\%$ edge and $26\%$ diagonal, aligned.

While \textit{ab-initio} MD is a powerful approach, the two principle limitations  
due to computational expense are 
finite size effects (as the simulation is infinitely periodic on $\approx$\si{nm}) and short timescale 
($\approx$\si{ps} are insufficient for diffusion processes). 
To access the time and length scales necessary to represent realistic
non-equilibrium structures, and so directly comparable to experiment, we
construct a classical model for molecular dipole interactions.

\textit{Polar molecules on a lattice.}
We start from the lattice dynamical theory of ferroelectricity
(see P.W. Anderson\cite{anderson_career_1994} and
W. Cochran\cite{cochran_crystal_1960,cochran_crystal_1961}.) 
We limit ourself to assuming that the dominant soft phonon mode in the system
is the free rotation of the molecular dipole within an extended perovskite
cage structure.  

The treatment of polarisation as an effect of rotational Brownian motion is
analytically challenging\cite{mcconnell_rotational_1980}.  
Here we simulate these physics numerically by using a Monte-Carlo method to
calculate the equilibrium configuration of the dipoles.
The macroscopic response of the material is expected to be linked to very 
slow rearrangements of domain walls as a result of statistically rare 
cooperative rearrangements of the microscopic dipoles. 
Therefore we need to integrate a long way in simulation time to 
equilibrate the material.

We construct a model Hamiltonian for the dipoles (vectors $p_i$) by summing the interaction
energy of the applied unshielded electrostatic field ($E_0$), near-neighbour dipole-dipole interactions and local
cage strain ($K$)
\begin{align}
\hat{H} = &\sum^n_{dipole,E-field} & (p_i.E_0) \\
+ &\sum^{n,m}_{dipole,dipole} &\frac{1}{4.\pi \epsilon_0} (\frac{p_i.p_j}{r^3}-\frac{3(\hat{n}.p_i)(\hat{n}.p_j)}{r^3}) \\
    + &\sum^n_{dipole,strain} &K.(|p_i.\hat{x}|+|p_i.\hat{y}|)
\end{align}  
Here the energies are calculated with numerically efficient dot products operating on Cartesian three-vectors. 
The unit vector $\hat{n}$ is constructed along the vector $r$ between the
dipoles under consideration, and $\hat{x},\hat{y}$ are unit lattice vectors. 
The Monte-Carlo method progresses with a Metropolis algorithm. 
A random lattice position is chosen, and a random new direction for the cation molecular dipole. 
The energy change ($\Delta E$) is computed 
with unshielded ($\epsilon_r =1$) dipole-dipole
interactions.  
Exothermic moves are automatically accepted; endothermic moves
are accepted if $\gamma < e^{-\beta \Delta E}$ where $\gamma$ is a random
variable on $[0,1]$, $\beta$ is $1/k_BT$.
For numerical efficiency we simulate on-lattice, with a cut-off for 
dipole-dipole interactions of three lattice units; this allows for $10^6$ 
attempted Monte-Carlo moves per second on standard hardware. 

At equilibrium, we associate an electric displacement $D$ related to $E_0$
and the polarisation density by
\begin{equation}
D =  E_0 + 4\pi P
\end{equation}
The polarisation density $P$ can be calculated by a summation over the
microscopic dipoles. 
$\epsilon_s$ refers to the static relative permittivity, rather than dielectric
constant, as it is a function of temperature.
\begin{equation}
\epsilon_s = 1+ \frac{4\pi P}{E_0}
\end{equation}
We can reconstruct the dipole potential felt at an arbitrary lattice site by
summing the potential contribution from all other lattice sites
\begin{equation}
    V^{dipole}_{i,j} = \frac{1}{4.\pi \epsilon_0}  \sum_{sites} \frac{p.r}{|r|^3}
\end{equation}
The main simulation variables are the strength of the interactions. 
Considering a point dipole-dipole interaction between unshielded
methyl-ammonium dipoles ($2.29$D)\cite{frost-2014} at the unit cell spacing of 6.29 \si{\angstrom}, this energy is $\approx25$ \si{\milli\electronvolt},
which we take in this work as exactly $1 k_B$T (T = 300\si{\kelvin}). 
From here we take $K=0$, allowing the dipoles to freely rotate without frustration. 

\textit{Equilibrium behaviour at room temperature.}
The domain structure of an equilibrated film with zero applied field is shown in Figure \ref{25x25_arrowdiagrams_300K_zerofield}. 
Twinning of the MA dipoles occurs to minimise the free energy. 
This leads to aligned domains along the square axes (as cofacial alignment
minimises the dipole-dipole distance compared to diagonal). 
Visualising the resulting dipole potential that comes from this alignment, we
observe the presence of structured interpenetrating regions of high and low
electric potential, following the features in the domain boundaries.  

The temperature dependence of the domain structure is shown in Figure \ref{10x10_eqm_dipoles_vs_T}. 
At 0 K, a striped anti-ferroelectric phase is favoured, which becomes increasingly 
disordered as the temperature increases due to the 
role of configuration entropy. 
The room temperature phase could be viewed as superparaelectric, 
consisting of randomly oriented linear ferroelectric domains, while the 
domains are broken to give a paraelectric phase at 1000 K.

The effective simulation temperature varies linearly with respect to the Hamiltonian interaction energies. 
As our model currently has a point-dipole approximated interaction energy, and
ignores energetic contributions from the cage strain, or ion inertia (freely
rotating dipoles), simulation temperature cannot be expected to correspond directly
to physical temperature.

\textit{Electric field dependence of polarisation.}
For a solar cell to operate the electrical contacts must be selective---a 
difference in work function must exist between the front and back contacts.
This selectivity induces a built-in potential that at short--circuit (or in the
dark) results in an electric field across the device, which acts to sweep out 
generated charge. 
Current collection at short--circuit is therefore generally maximised. 
Here we assume that the perovskite solar cells, without intentional doping, 
are $p-i-n$ type; the potential drops linearly across the active material, producing a constant electric field. 
When a voltage is applied in forward bias, it works to counteract this built-in field.
At the maximal power point for a relatively optimal solar cell material such as
MAPI, the operating voltage is close to the open--circuit voltage, which is
close to the built-in potential and so only a small electrical field will apply
across the device.

In typical perovskite solar cells, the layer thickness is of order hundreds of
nanometres, and the built in voltage of  $\approx1$ \si{\volt}.  
In the absence of charge equalisation effects, this results in an 
electric field of 1 to 10 \si{\mega\volt\per\metre} across the hybrid perovskite. 
%
The interaction energy of the MA dipole with this field is
$U=-p.E$, 0.48 \si{\milli\electronvolt} for a upper limit field strength of 
10  \si{\mega\volt\per\metre}. 
This is a relatively small perturbation compared to the dipole-dipole
interaction of $25$ \si{\milli\electronvolt}. 

The alignment of the dipoles as a result of the applied field is shown in
Figure \ref{25x25_field_fourier}; the dipoles rearrange to partially counteract
the applied field.  
The response in overall dipole alignment is $0.5$\si{\percent} in the direction of the field 
(versus $0.04$\si{\percent} background fluctuation for no-applied field 
We emphasise that it took $10^4$ Monte-Carlo moves per lattice site to achieve
this equilibrium structure, well beyond where total energy and Debye
polarisation appeared to have approached their equilibrium asymptotes.

Though such a small perturbation from the built-in field exhibits negligible visual 
effect on the alignment (Figure \ref{25x25_field_fourier} -- Top), the effect on
the dipole potential (Figure \ref{25x25_field_fourier} -- Middle) is strong,
leading to deeper more segregated regions of positive and negative potential. 

We quantitatively evaluate the change in the structure of the dipole potential
by a two-dimensional Fourier transform (Figure \ref{25x25_field_fourier} -- Bottom). 
Here we see that the zero field (open--circuit) equilibrium structure is equally
distributed in both axes, and the density along the origin indicating the
presence of linear features. 
%
%
We interpret these short--circuit features as being the development of carrier
traps by the dipole domain response to the built-in field; at
open--circuit the extended ridges and valleys in potential could act as
channels in which charge transport can take place. 
Understanding the true role of these features and quantifying the effect on
device performance will require a sophisticated device model and an improved
understanding of the nature of charge carriers and charge transport in this class of materials. 

Particularly, the size of the carrier polarons in these materials will heavily affect
the influence of such local inhomogeneiity in the electrostatic potential on
charge transport.
The Fr\"ohlich coupling constant is estimated to be $\alpha=1.2$ using published band
parameters for MAPI ($m^*=0.12$, $\epsilon_0=24.1$, $\epsilon_\infty=4.5$) and
a longitudinal optical phonon frequency of 9 THz. 
This indicates an intermediate
electron-lattice coupling, phonon dressing leading to a 25\% increase in the
effective mass by the Feynman variational treatment\cite{feynman_slow_1955}.
The electron and hole polaron radii would correspond to approximately
5 perovskite unit cells, sufficiently small to be influenced by 
inhomogeneity in the local electrostatic potential we predict. 
We therefore consider it plausible that such variations in
electrostatic potential will drive both carrier polaron segregation (open--circuit 
dipole structure) and trapping (short--circuit dipole structure),
leading to increased bulk recombination at short--circuit.

Recently we have been made aware of inverted MAPI perovskite cell designs
capped with a fullerene electron accepting layer, which exhibit 
reduced hysteresis.\cite{xiao2014efficient}
We interpret this as a result of gaps in the MAPI film resulting in
high penetration of the fullerene, forming a heterojunction. 
This effectively quenches the recombination, as electron extraction into the
fullerene phase out competes recombination.
In comparison to fullerene, \ce{TiO2} is a poor (slow) electron acceptor.

\textit{Additional ferroelectric contributions.}
Beyond molecular dipole reorientations, additional ferroelectric contributions
include distribution of free carriers (electrons and holes),
as well as rotations and titling of the \ce{PbI3-} cage structure. 
The formation of points defects including 
\ce{I2-} defect complexes\cite{du-2014} and charged Pb vacancies\cite{yin-063903}
have been suggested, which may also respond to an applied electric field. 
Previous impedance analysis have also suggested room temperature ionic
conductivity in these materials.\cite{maeda1997dielectric}

Under short--circuit conditions the MAPI layer will be polarized due to the
alignment of dipoles, as demonstrated in our simulations. 
If the system is then placed in open--circuit conditions the polarization of
MAPI is removed; however, the depolarization field consisting of charge carriers
will take time to re-equilibrate.
There will necessarily be feedback between the 
ferroelectric domain structure (slower process) and carrier distribution (faster process), 
therefore causing a further hysteretic contribution to current-voltage measurements. 
Additionally, the electric field across the absorber (under short--circuit conditions) 
would lead to an alignment
of the cage polarisation (rotation and titling) and domain structure present in the hybrid perovskite film,
which is likely to be over a longer time scale than the dipole reorientation. 
The realignment of these domains upon removal or reversal of the field is
another possible source of hysteresis. 

In this work we have shown through \textit{ab-initio} molecular dynamics that the
methyl-ammonium ion is rotationally mobile in hybrid perovskites at room temperature, 
and that the material is structurally soft. 
This material behaviour may be fortuitous in terms of facilitating transport
across grain boundaries when combined with the calculated large polaron
transport and small effective masses.\cite{frost-2014} 
The large site variation of the ions deserves further study in terms of its
effect on material polarisation and ferroelectric response.
%
Further investigation of the molecular dynamics will include expanding the
simulation volume and analysing the trajectories further.  
Additional work with the Monte-Carlo codes are required to extend the
simulation to three dimensional perovskite volumes, introduce other move types
(such as movement of ions both within the lattice cells, and as net migration
through the film), and extend the simulated experiments to other relevant
device physics.
More sophisticated interaction terms for the molecular cations would increase 
the expected accuracy of these codes. 

In conclusion, we have investigated the behaviour of the dipolar
methyl-ammonium cation in \ce{CH3NH3PbI3} using numerical simulations, which
have provided insights into the  domain structure and polarisation fields,
which will be important for developing quantitative models
to explain the unusual device physics of hybrid perovskite solar cells.

\textit{Acknowledgements.}
We are grateful for useful discussions with Piers Barnes and Aur\'elien Leguy (Imperial College London), chiefly concerning the role of the electric field in these materials. 
The molecular dynamics in this work was instigated to interpret their (unpublished) neutron scattering data. 
We thank Laurie Peter and Petra Cameron (University of Bath) for useful discussions on hysteresis in perovskite solar cells. 
We acknowledge membership of the UK's HPC Materials Chemistry Consortium, which is funded by EPSRC grant EP/F067496. 
Additional computing resources were provided via the PRACE project UltraFOx. 
J.M.F. and K.T.B. are funded by EPSRC Grants EP/K016288/1 and EP/J017361/1, respectively.
A.W. acknowledges support from the Royal Society and ERC (Grant 277757). 

~

\textbf{Computational Details}

The set-up for density functional theory (DFT) calculations of the primitive
unit cells of \ce{CH3NH3PbI3} with a range of molecular orientations, including
structure optimisation and static dielectric
response,\cite{brivio-042111,brivio-155204} as well as the lattice polarisation
and barriers to rotation,\cite{frost-2014} have been previously reported.
These were taken as the starting point for the MD simulations in this study. 

\textit{Molecular dynamics (MD).}
Finite temperature Newtonian MD simulations were performed based on the atomic
forces calculated at each timestep using DFT ($\Gamma$-point sampling with the
\textsc{PBESol} functional and a 400 eV plane wave cutoff).  
The starting configuration was a fully relaxed $2\times2\times2$ pseudo-cubic
supercell (with a $3\times3\times3$ $k$-grid) with MA ions aligned along the
$<100>$ direction. 
Spin-orbit coupling is not treated primarily due to the prohibitive
computational expense; as orbital occupation is not changed for the undoped system, the
effect on the atomic forces is expected to be negligible.  
Trajectory data were collected every 50 integration steps (25 \si{fs}). 
A Nos\'{e} thermostat (canonical ensemble) was used with a Nos\'{e} mass of 3.
Custom codes were written for the analysis, with the help of the \textsc{MDAnalysis} library.\cite{mdanalysis_2011} 
A total of 58 \si{ps} (2319 frames) of data was used for analysis, after an equilibration run of 5 \si{ps}. 
This generated 18547 unique MA alignment vectors.

\textit{Monte Carlo (MC).}
The MC implementation uses a Mersenne
Twister\cite{matsumoto_mersenne_1998} pseudo random number generator; the code
is serial (running at $10^6$\si{\per\second} on modest hardware), but efficient uses of computational resources is achieved by making
ensemble runs with GNU Parallel\cite{Tange2011a}. 
The modestly sized simulations presented here can be extended up to full device sized simulations, with well defined statistics over ensemble runs. 
The initial state is a lattice of randomised dipoles. 
The resulting classical Hamiltonian dipole Monte Carlo code, \textsc{StarryNight}, is available as a source code repository on GitHub.\cite{GitHub_StarryNight} 
Codes to interpret the \textit{ab-initio} molecular dynamics used in the production of this manuscript are similarly available.\cite{GitHub_MAPI-MD-analysis}

\bibliography{library}

\newpage

\begin{figure*}[ht]
    \centering
    \begin{subfigure}[b]{0.5\textwidth}
        \includegraphics[width=\textwidth]{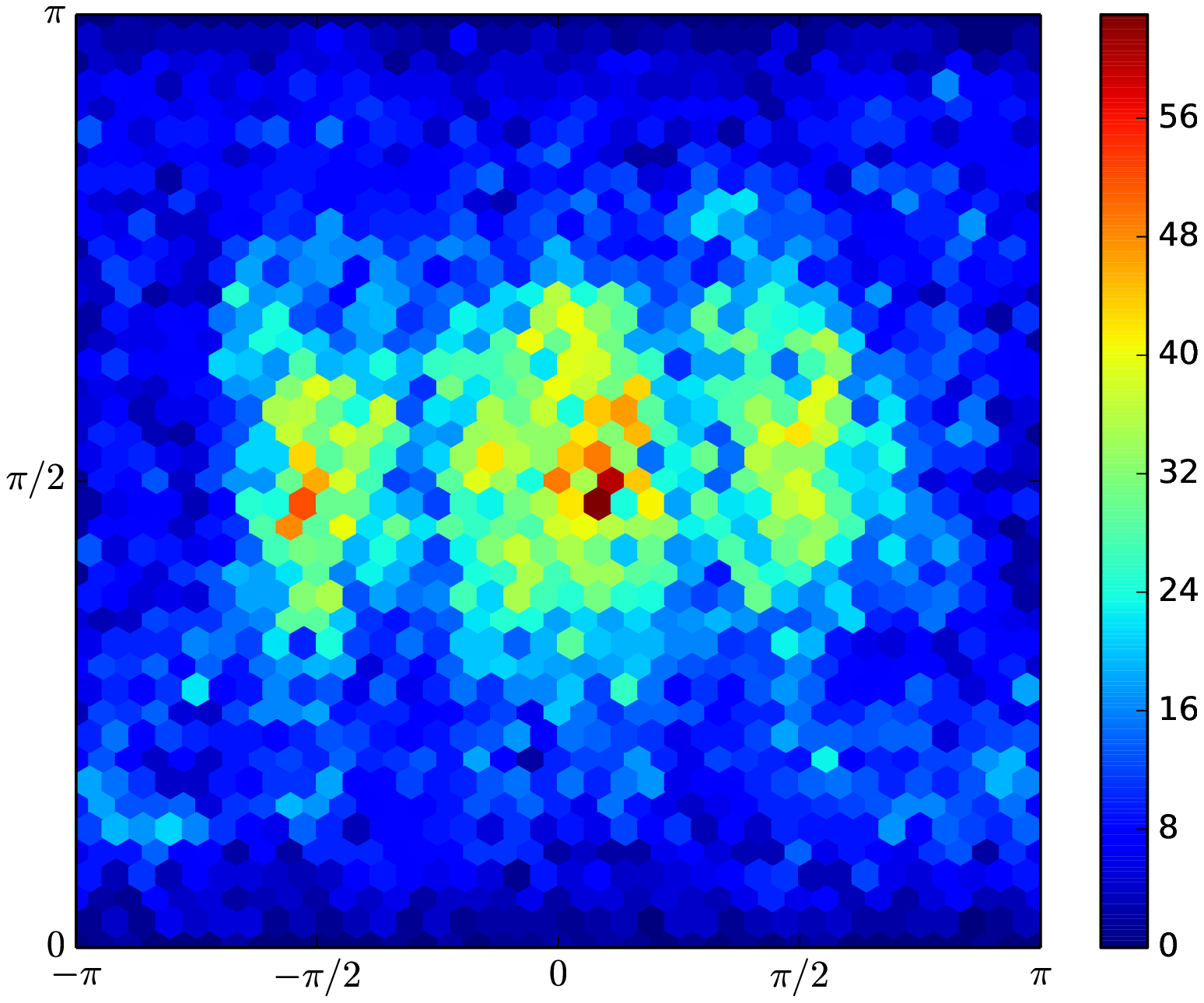}
        \caption{Spherical coordinates, no symmetry}
        \label{theta_phi_nosymm}
    \end{subfigure}%
    ~ 
    \begin{subfigure}[b]{0.5\textwidth}
         \includegraphics[width=\textwidth]{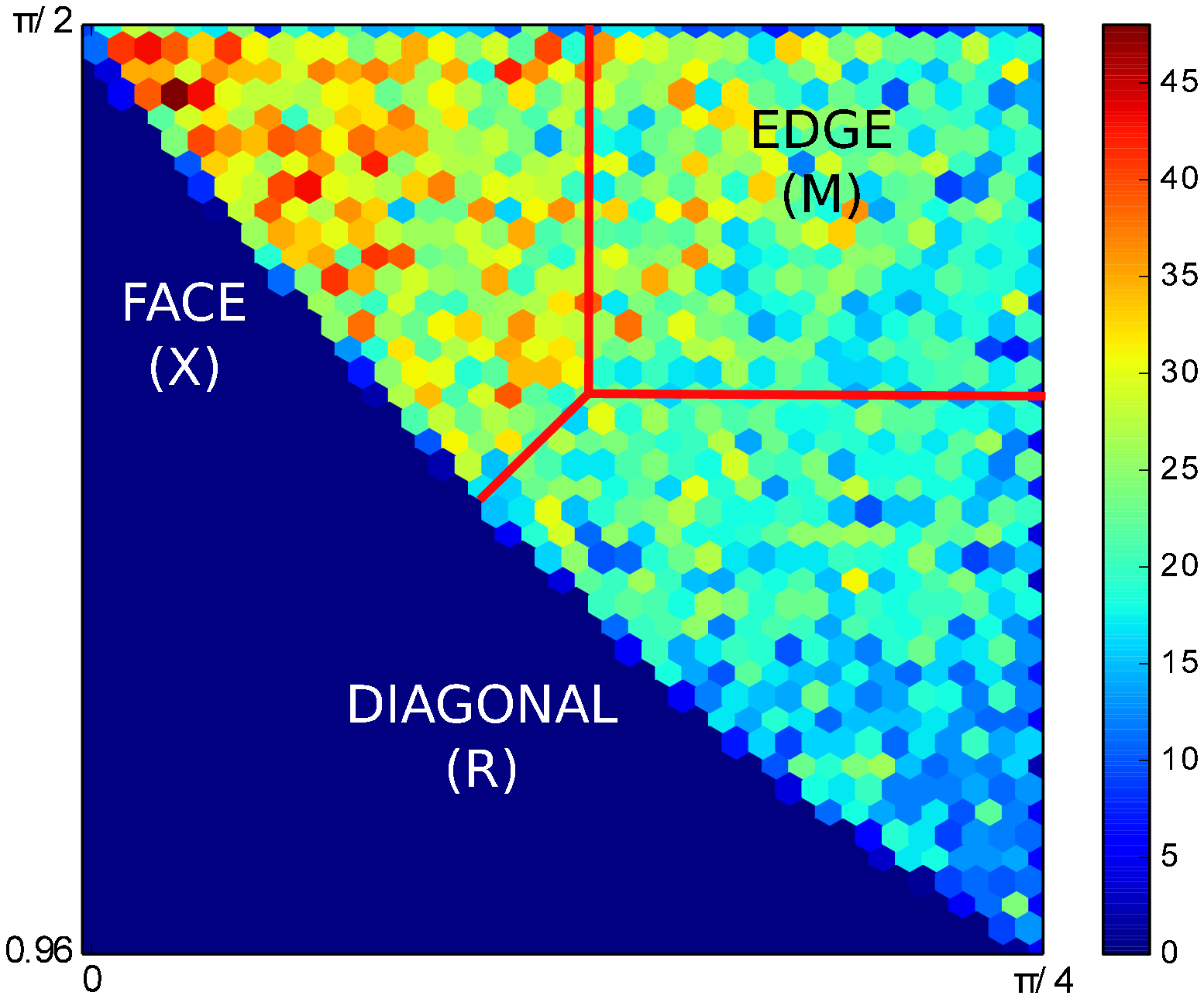}
        \caption{Full 48-fold ($O_h$) symmetry}
        \label{theta_phi_symm}
    \end{subfigure}
    \caption{Density maps (2D histograms, in spherical coordinates) of MA alignment within the pervoskite cage
    structure as determined by \textit{ab initio} molecular dynamics at 300 K.
        The data are centered on $\phi,\theta=0$ being facial orientation. 
        The symmetry folded data bounds the segment between diagonally aligned in the cube (bottom right), pointing at an edge (top right) and pointing at a face (top left). 
        This covers a segment of the original data bounded by $0<\theta<\pi/4$ and $arcos(\frac{1}{\sqrt{3}}) < \phi < \pi/2$.
        In total 18547 data points (8 MA alignments per data frame) are hexagonally binned, with 36 bins in the x-axis.
} 
    \label{theta_phi}
\end{figure*}

\begin{figure*}[h!]
\centering
\begin{subfigure}[b]{0.5\textwidth}
    \includegraphics[width=\textwidth]{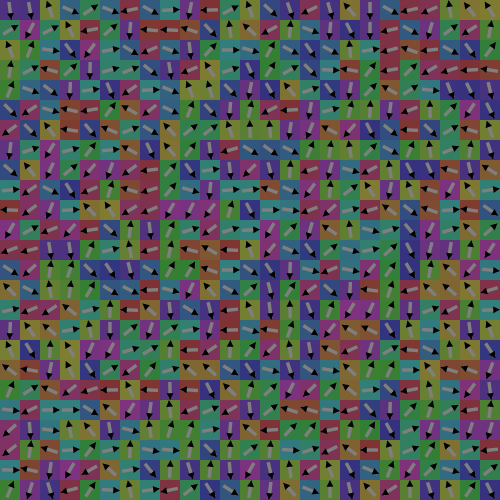}
\end{subfigure}%
~ 
\begin{subfigure}[b]{0.5\textwidth}
    \includegraphics[width=\textwidth]{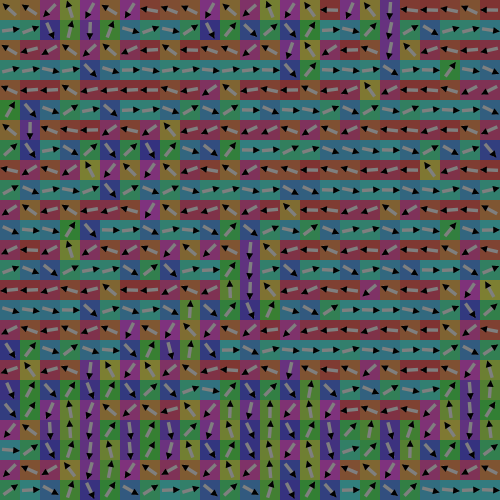}
\end{subfigure}%

\begin{subfigure}[b]{0.5\textwidth}
    \includegraphics[width=\textwidth]{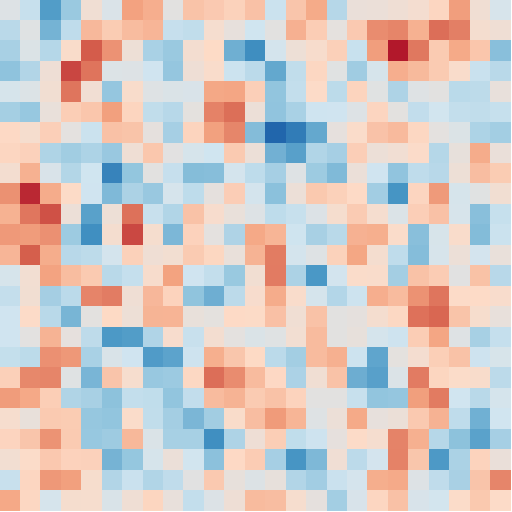}
    \caption{Random starting configuration}
\end{subfigure}%
~ 
\begin{subfigure}[b]{0.5\textwidth}
    \includegraphics[width=\textwidth]{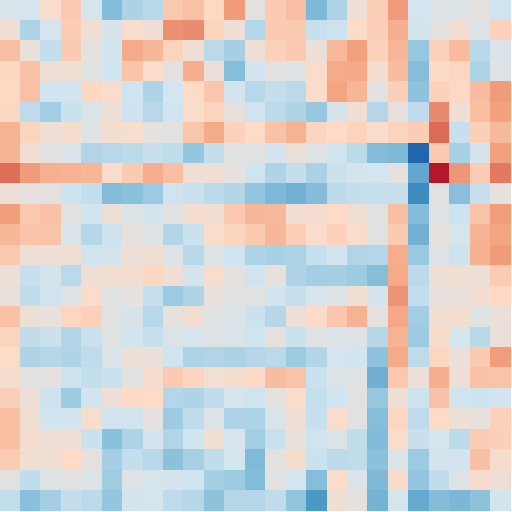}
    \caption{After equilibriation}
\end{subfigure}%

\caption{(Top) Representative orientation of dipoles in a 25$\times$25 two-dimensional periodic slab representing \ce{CH3NH3PbI3}, initial random configuration (left) and after equilibration under Monte Carlo at 300 K (right). (Bottom) Electrostatic potential as a result of alignment of dipoles, calculated by summation over all other lattice sites for each site.
Note the emergence of twinned domain structure formed of linear dipole chains. 
In the dipole potential these give rise to structured interpenetrating regions of positive and negative electric potential.  
}
\label{25x25_arrowdiagrams_300K_zerofield}
\end{figure*}

\begin{figure*}[ht!]
\begin{subfigure}[b]{0.25\textwidth}
    \includegraphics[width=\textwidth]{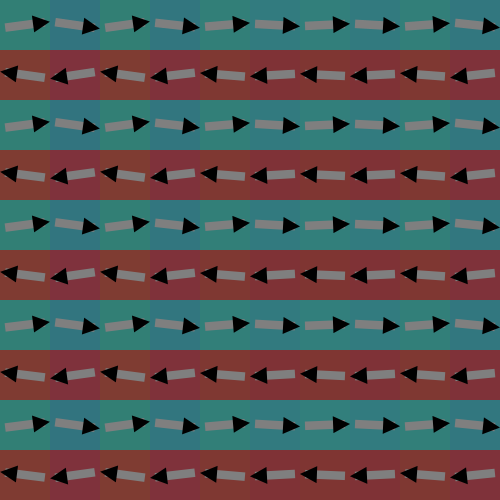}
    \caption{0\si{\kelvin}}
\end{subfigure}%
~ 
\begin{subfigure}[b]{0.25\textwidth}
    \includegraphics[width=\textwidth]{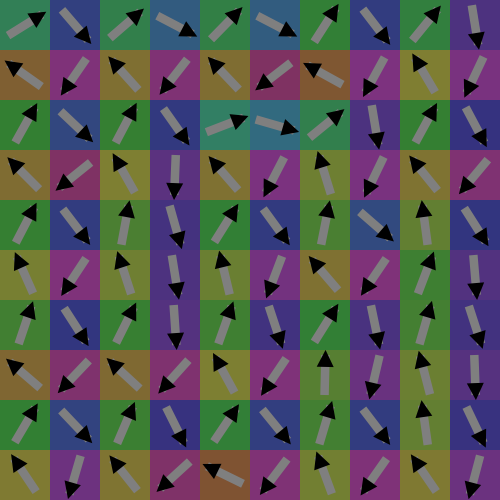}
    \caption{100\si{\kelvin}}
\end{subfigure}%
~ 
\begin{subfigure}[b]{0.25\textwidth}
    \includegraphics[width=\textwidth]{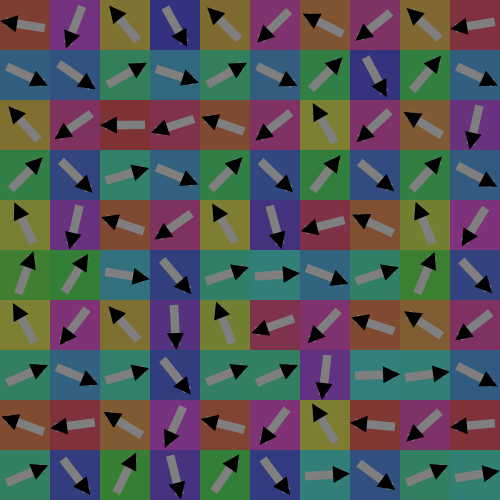}
    \caption{300\si{\kelvin}}
\end{subfigure}%
~ 
\begin{subfigure}[b]{0.25\textwidth}
    \includegraphics[width=\textwidth]{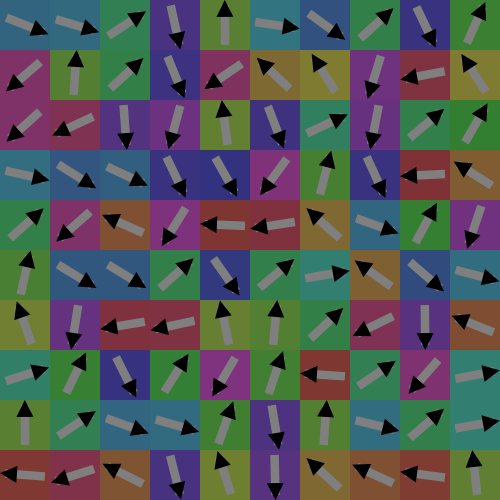}
    \caption{1000\si{\kelvin}}
\end{subfigure}%
\caption{
Methyl-ammonium dipole alignment as a function of temperature. 
Low temperature results in complete alignment and large domains, which disorder with increasing temperature leading to small domains, eventually resulting in completely disordered dipoles in the high temperature limit. 
The physical temperature of these transitions is difficult to quantify, being linear in the dipole-dipole interaction potential (here 25 \si{\milli\electronvolt}).}
\label{10x10_eqm_dipoles_vs_T}
\end{figure*}

\begin{figure*}[ht!]
\begin{subfigure}[b]{0.38\textwidth}
    \includegraphics[width=\textwidth]{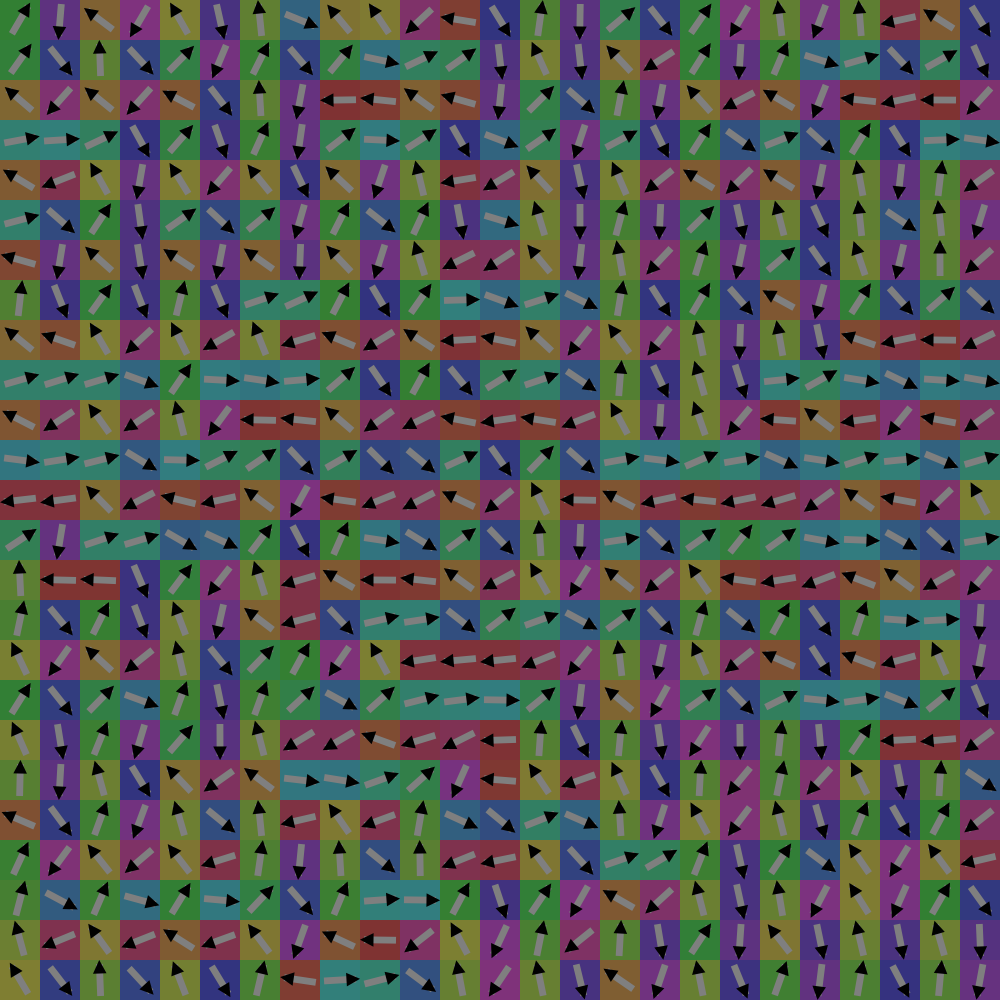}
\end{subfigure}%
~ 
\begin{subfigure}[b]{0.38\textwidth}
    \includegraphics[width=\textwidth]{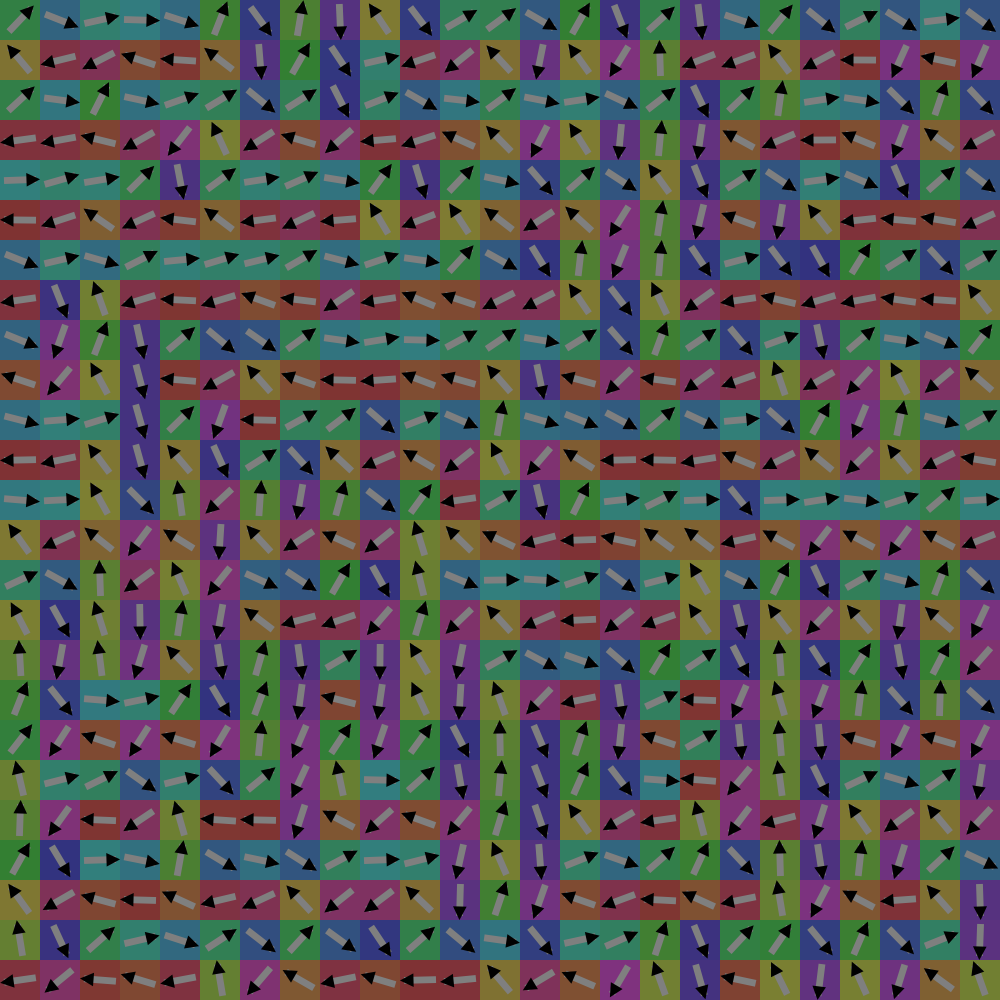}
\end{subfigure}

\begin{subfigure}[b]{0.38\textwidth}
    \includegraphics[width=\textwidth]{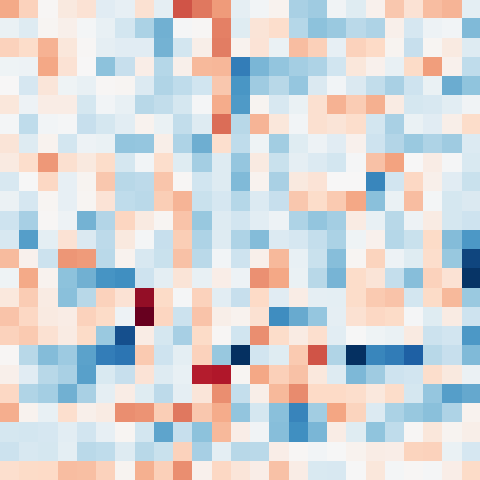}
\end{subfigure}%
~ 
\begin{subfigure}[b]{0.38\textwidth}
    \includegraphics[width=\textwidth]{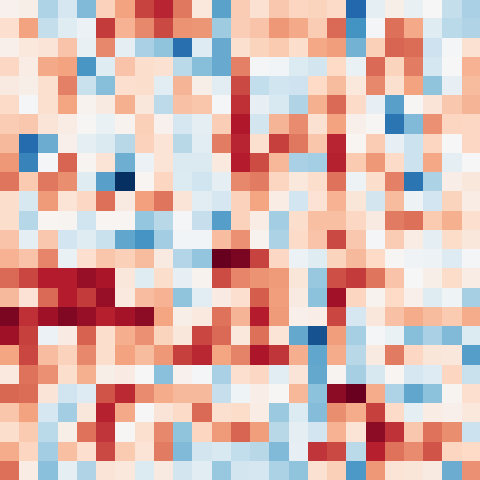}
\end{subfigure}

\begin{subfigure}[b]{0.38\textwidth}
    \includegraphics[width=\textwidth]{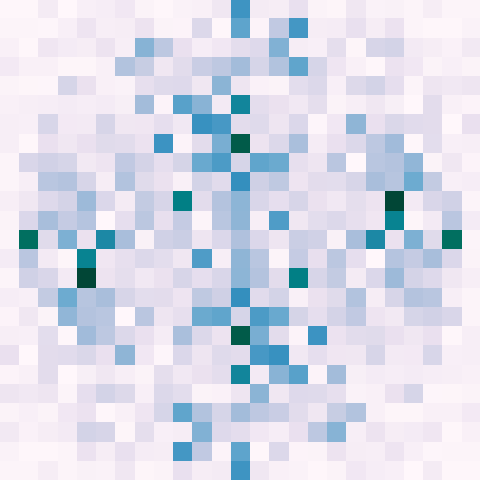}
    \caption{Open--circuit (no applied field)}
\end{subfigure}%
~ 
\begin{subfigure}[b]{0.38\textwidth}
    \includegraphics[width=\textwidth]{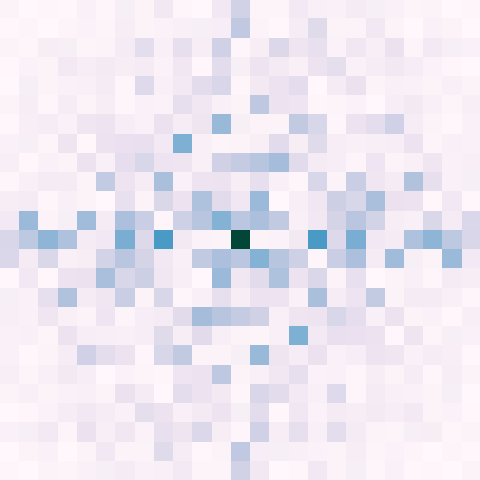}
    \caption{Short--circuit (field 10 \si{\mega\volt\per\metre})}
\end{subfigure}

\caption{Analysis of the dipole alignment in a 25$\times$25 \ce{CH3NH3PbI3} pervoskite film, equilibrated at open and short--circuit conditions. (Top) Alignment of the dipoles, (Middle) dipole potential, (Bottom) 2D Fourier transform of dipole potential (zero frequency component shifted to the centre), showing the change in periodicity of the structure.}
\label{25x25_field_fourier}
\end{figure*}

\clearpage

\begin{center}
\textbf{\large Supplemental Materials: Ferroelectric contribution to anomalous hysteresis in hybrid perovskite solar cells}
\end{center}
\setcounter{equation}{0}
\setcounter{figure}{0}
\setcounter{table}{0}
\setcounter{page}{1}
\makeatletter
\renewcommand{\theequation}{S\arabic{equation}}
\renewcommand{\thefigure}{S\arabic{figure}}
\renewcommand{\bibnumfmt}[1]{[S#1]}
\renewcommand{\citenumfont}[1]{S#1}

\maketitle

\begin{figure*}[ht]
    \centering
    \begin{subfigure}[b]{0.5\textwidth}
        \includegraphics[width=\textwidth]{figures_hexbin_figures_no-symm_fullaxes-mdanalysis_cn_dist.eps}
        \caption{No symmetry}
    \end{subfigure}%
    ~ 
    \begin{subfigure}[b]{0.5\textwidth}
        \includegraphics[width=\textwidth]{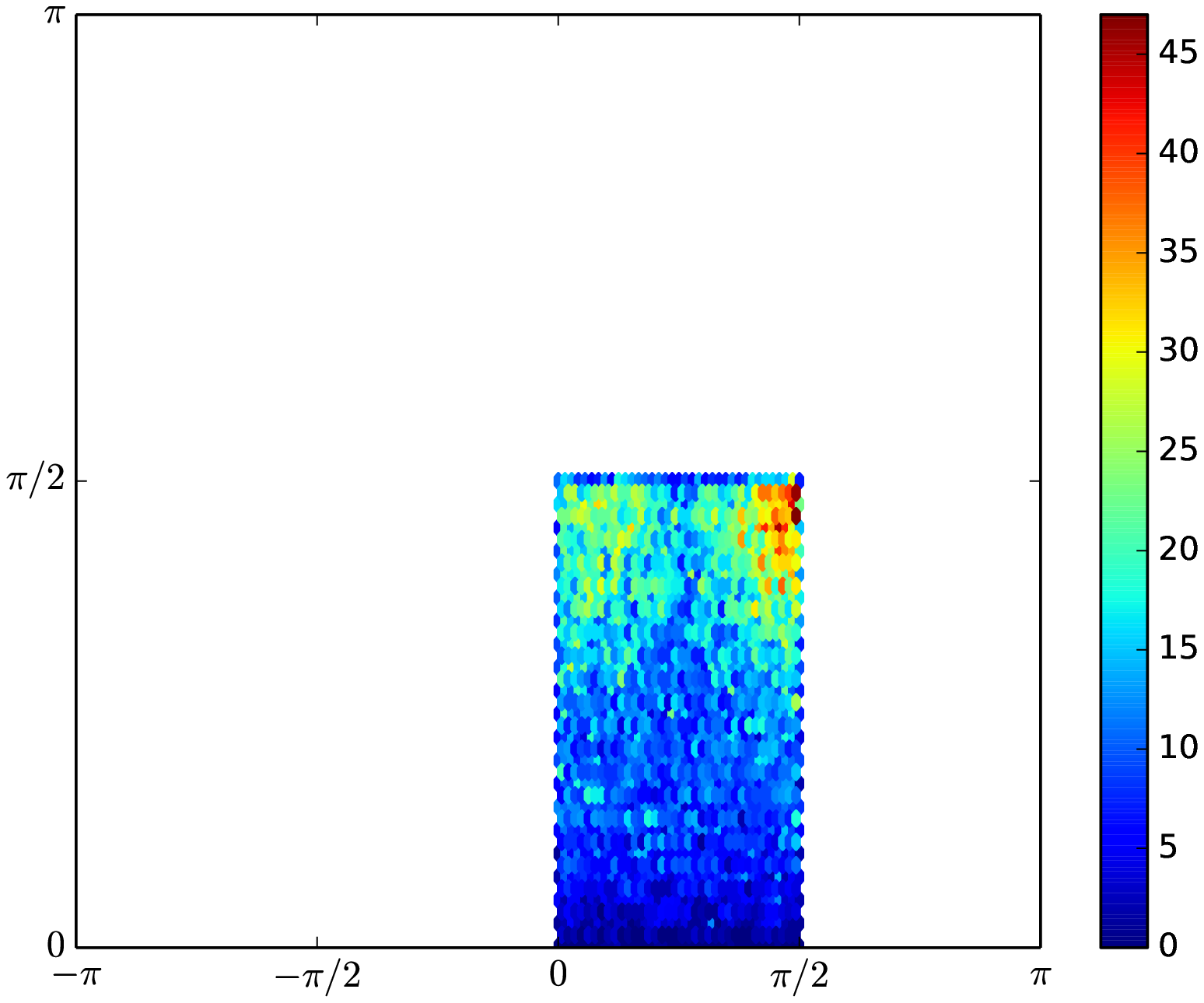}
        \caption{Octant, 8-fold symmetry}
    \end{subfigure}
    ~ 
    \begin{subfigure}[b]{0.5\textwidth}
        \includegraphics[width=\textwidth]{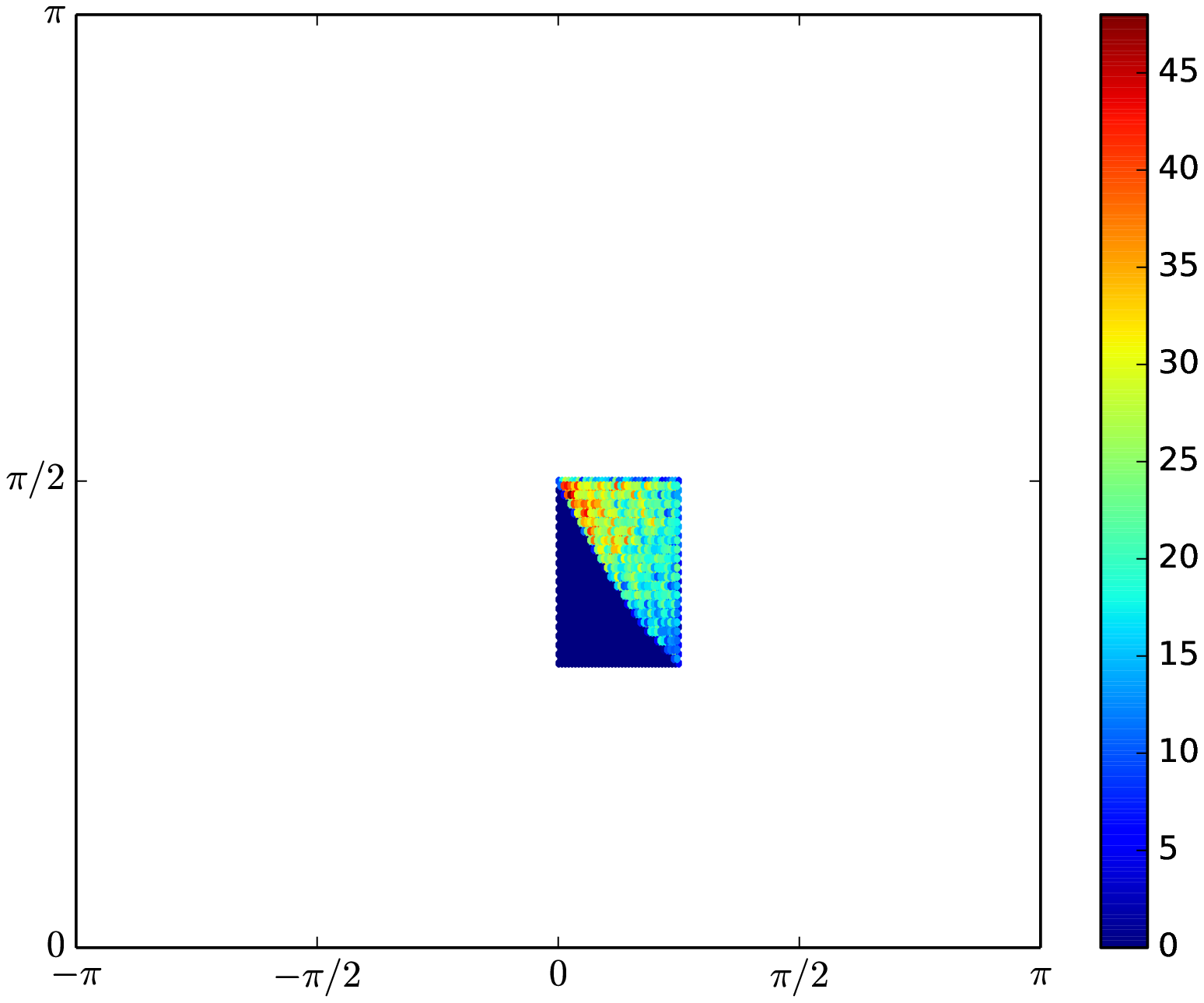}
        \caption{Full 48-fold symmetry}
    \end{subfigure}

    \caption{Density maps (2D histograms, in spherical coordinates) of MA alignment within the pervoskite cage.
        The data are centered on $\phi,\theta=0$ being facial orientation.
    Here we present the higher symmetry reduced data on the same full $-\pi<\theta<\pi$, $0<\phi<\pi$ spherical coordinate axes.}
    \label{SI:theta_phi}
\end{figure*}

Exploitation of the $O_h$ symmetry of the cubic perovskite lattice is essential
to improve the signal-to-noise ratio of computationally expensive
molecular dynamics data.
However this is difficult to visualise, and so we provide Figure
\ref{SI:theta_phi} explicitly showing how starting (a) with no symmetry, we first
(b) fold the data into the first octant and then reflect along the three lines
of symmetry radiating from the diagonal to arrive at (c), a 48 fold increase in
density.

To reduce an arbitrary three-vector onto this reflection domain, we first take
the absolute value of the Cartesian components (reflecting into the first
octant), then sort the elements of the vector by size (which applies
a three-plane reflection about $[1,1,1]$).
This is implemented in \textsc{Python} as \texttt{sorted(abs(r))}.

\begin{figure}[ht]
    \centering
    \includegraphics[width=\columnwidth]{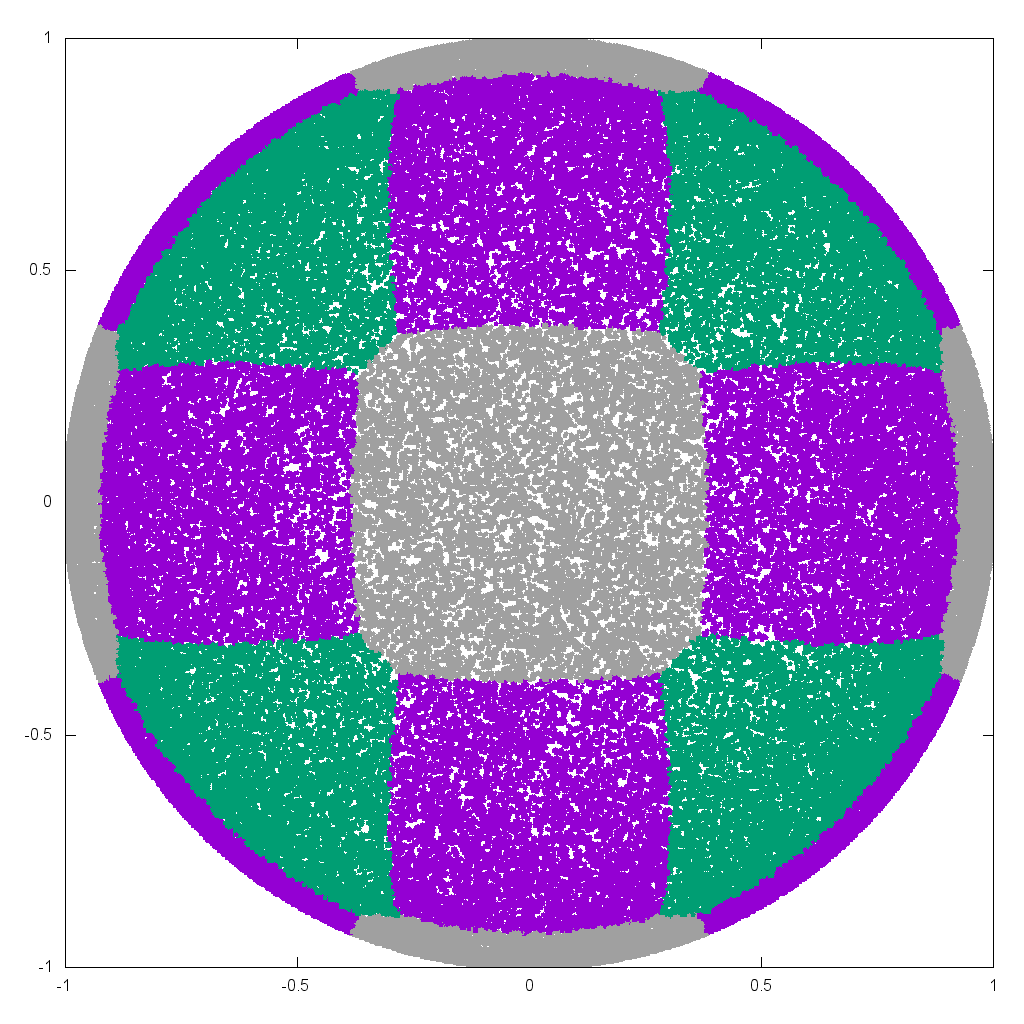}
    \caption{Projection onto the x,y plane of the $10^5$ random spherical
    points used to weight the population of alignments of MA ions. Colour represents which of face (grey), edge (purple) and diagonal (green) vectors the sphere point is closest in angle to. Raw populations from this integration were $[27312, 44335, 28353]$, respectively.}
    \label{SI:spherical_coordinate_collapse}
\end{figure}

In this work we alternatively define the histogram bins for our methylammonium
vectors by their angular nearness to the face ($[1,0,0]$), edge ($[1,1,0]$) and
diagonal ($[1,1,1]$) vectors, with respect to the cubic perovskite lattice.
This partitions the cube into 6 regions around the face, 12 regions around the
edges and 8 regions around the diagonals.
The boundary between these regions is midway on the grand-circle route between
the high symmetry points; this is similar in construction to a Wigner-Seitz
cell but on the surface of a sphere.
In Figure \ref{SI:spherical_coordinate_collapse} we present a projection of the
$10^5$ random spherical vectors used in Monte-Carlo integration to define the
area of these zones of influence and thus the weighting to allow direct
comparison of methylammonium alignment.

\end{document}